\documentclass{ws-p9-75x6-50}

\begin{document}

\title{Coulomb gap in the quantum Hall insulator}

\author{Michael Backhaus and Bodo Huckestein}

\address{Institut f\"ur theoretische Physik, Universit\"at zu K\"oln,
  D-50937 K\"oln, Germany\\E-mail: bh@thp.uni-koeln.de}


\maketitle

\abstracts{
We calculate numerically the spectrum of disordered electrons in the
lowest Landau level at filling factor 1/5 using the self-consistent
Hartree-Fock approximation for systems containing up to 400 flux
quanta. Special attention is paid to the correct
treatment of the $q=0$ component of the Coulomb interaction. For
sufficiently strong disorder, the system is an insulator at this
filling factor. We observe numerically a Coulomb gap in the
single-particle density of states (DOS). The DOS agrees quantitatively 
with the predictions for classical point charges.
}


The Coulomb interaction between charged localized particles leads to a
vanishing single particle density of states (DOS) at the Fermi energy,
$\rho(E)$. This classical effect was understood by Efros and
Shklovskii as a necessary condition for the stability of the ground
state with respect to particle-hole excitations \cite{ES75}. A more
quantitative description of this Coulomb gap was obtained by Efros
using a self-consistent method \cite{Efr76}.

The argument leading to the Coulomb gap is expected to hold in a
quantum mechanical system, too, if the localization length of the
particles is small compared to their distance. An important example of
such a system is a disordered two-dimensional electron gas in a strong
magnetic field, exhibiting the integer quantum Hall effect
\cite{KDP80}. In this system, all states are exponentially localized
except those at single energies close to the centers of each Landau
level \cite{Huc95r}. Indeed, Yang and MacDonald numerically found a
linearly vanishing DOS when the filling factor $\nu$ of the lowest
Landau level was 1/5 \cite{YM93}. Remarkably, the vanishing of the DOS
persists even at the critical energy in the band center
\cite{YM93,HB99}.

In this communication, we show that the DOS in the lowest Landau level
at filling factor 1/5 can be quantitatively described by the
self-consistent method of Efros \cite{Efr76}. We carefully consider
the finite-size effects in the self-consistent Hartree-Fock (HF)
approximation and apply Efros' ideas to the finite systems considered
in the numerical simulations. The agreement is surprising, since the
effects of wavefunction overlap and exchange interaction are present
in the distribution of the HF matrix elements.


Before restricting our discussion to the QH system, we want to discuss 
the self-consistent Hartree-Fock approximation of spinless (or fully
spinpolarized, as in the lowest Landau level) electrons on a finite
torus of area $L\times L$. The Hamiltonian is
\begin{equation}
  {\cal H} = \sum_{i=1}^{N} \left[T_i +
    U_{\mathrm{dis}}(\hat{\mathbf{r}}_i) + 
    V_{\mathrm{el-bg}}(\hat{\mathbf{r}}_i)\right] +
  \frac{1}{2}\sum_{\stackrel{\scriptstyle{i,j=1}}{i\neq j}}^N
  V(\hat{\mathbf{r}}_i-\hat{\mathbf{r}}_j) + V_{\mathrm{bg-bg}},
\end{equation}
where $T_i$ is the kinetic energy, $U_{\mathrm{dis}}$ the disorder
potential, $V_{\mathrm{el-bg}}$ the interaction with the homogeneous
background charge, $V$ the Coulomb interaction, and
$V_{\mathrm{bg-bg}}$ the background self-interaction,
respectively. The Hartree-Fock equations then take the form
\begin{equation}
  (T + U_{\mathrm{dis}} + V_{\mathrm{el-bg}}) \psi_\alpha(\mathbf{r}) +
  \sum_{\beta}f_\beta\sum_{\mathbf{q}}
  \left[(\beta|\beta)_{\mathbf{q}} \psi_\alpha(\mathbf{r}) - 
    (\beta|\alpha)_{\mathbf{q}} \psi_\beta(\mathbf{r})
  \right] v_q e^{iqr} = \varepsilon_\alpha\psi_\alpha(\mathbf{r}),
\end{equation}
with $(\beta|\alpha)_{\mathbf{q}} =
\langle\beta|e^{-i\mathbf{q}\cdot\mathbf{r}}|\alpha\rangle$, $v_q =
2\pi e^2/(L^2q)$ is the two-dimensional Fourier transform of the Coulomb 
interaction, and $f_\alpha$ is the Fermi function. The
$\mathbf{q}=\mathbf{0}$ part of the sum over $\mathbf{q}$ gives
\begin{equation}
  \label{q=0back}
  v_0\sum_{\beta}f_\beta
  \psi_\alpha(\mathbf{r})(1-\delta_{\alpha,\beta})=
  v_0(N-f_\alpha)\psi_\alpha(\mathbf{r})=
  -V_{\mathrm{el-bg}}\psi_\alpha(\mathbf{r})- 
  v_0f_\alpha\psi_\alpha(\mathbf{r}),
\end{equation}
which cancels the interaction with the background
charge up to a contribution proportional to $v_0=\int\!d^2r\,r^{-1}L^{-2}\sim
L^{-1}<\infty$. Since this contribution vanishes in the limit of
infinite system size it is often neglected resulting in the
HF equations
\begin{equation}
\label{eq:1}
  (T + U_{\mathrm{dis}}) \psi_\alpha(\mathbf{r}) +
  \sum_{\beta,\,\mathbf{q}\neq\mathbf{0}} f_\beta
  \left[(\beta|\beta)_{\mathbf{q}} \psi_\alpha(\mathbf{r}) - 
    (\beta|\alpha)_{\mathbf{q}} \psi_\beta(\mathbf{r})
  \right] v_q e^{iqr} = \varepsilon_\alpha\psi_\alpha(\mathbf{r}).
\end{equation}
However, the correct eigenvalues for the finite system would be
\begin{equation}
\label{eq:2}
  \tilde{\varepsilon}_i = \varepsilon_i - f_i v_{0},
\end{equation}
as an electron in an empty state interacts with $N$ other electrons,
while an electron in an occupied state interacts with only $N-1$ other 
electrons.


Efros based his self-consistent method \cite{Efr76} on the observation
that the ground state of a system is stable against every
particle-hole excitation. For the solutions $\varepsilon_i$ of
eq.~(\ref{eq:1}) this is equivalent to the condition
\begin{equation}
  \label{eq:3}
  \Delta E = \varepsilon_j -\varepsilon_i - v_{ij} > 0,
\end{equation}
for every $i,j$. In the classical limit, for states localized in an
infinitesimal vicinity of the position $\mathbf{r}_i$, the interaction
matrix element is given by
\begin{equation}
  v_{ij} = \sum_{\mathbf{q}\neq\mathbf{0}} v_q\,
  (i|i)_{\mathbf{q}} (j|j)_{-\mathbf{q}} =
  \frac{e^2}{|\mathbf{r}_i-\mathbf{r}_j|}-v_{0}.
\end{equation}
For the torus geometry,
\begin{equation}
  v_{0} = \frac{e^2}{L^2}\int_{L\times L} d^2r \frac{1}{r} 
  = 2 \ln (1+\sqrt{2})\frac{e^2}{L} \approx 1.76\frac{e^2}{L}.
\end{equation}

In the classical limit, eq.~(\ref{eq:3}) restricts the possible
distances $|\mathbf{r}_i-\mathbf{r}_j|$ and energies $\varepsilon_j$
for a given energy $\varepsilon_i$. This leads to a self-consistency
condition for the DOS \cite{Efr76}
\begin{equation}
  \label{eq:4}
  {\rho(\varepsilon)=\rho_0\exp\left[-{\pi\over
        2}e^4\int_0^{\infty}
      \frac{\rho(\varepsilon')\,d\varepsilon'}
      {(\varepsilon+\varepsilon'+{v_{0}})^2}\right]}.
\end{equation}
In contrast to the infinite system result of Efros, eq.~(\ref{eq:4})
in a finite system contains the finite contribution $v_0$.

In the limit of small $\varepsilon$ and large $L$, eq.~(\ref{eq:4})
has the asymptotic solution
\begin{equation}
  \label{eq:6}
  {\rho(\varepsilon) = \frac{2}{\pi e^4} ({v_{0}} + |\varepsilon|)}.
\end{equation}

The stability condition for the true HF eigenvalues
$\tilde{\varepsilon_i}$ (eq.~(\ref{eq:2})) is
\begin{equation}
  \Delta E = \tilde{\varepsilon}_j -\tilde{\varepsilon}_i - v_{ij}
  - v_{0} > 0.
\end{equation}
The self-consistency equation for the density of the $\varepsilon$
does not contain $v_0$. However, now a lower bound for the energy
difference $\tilde{\varepsilon}_j -\tilde{\varepsilon}_i$ exist, since 
the distance between the states can be at most $L/\sqrt{2}$ on the
torus:
\begin{equation}
  \label{eq:5}
  {\rho(\tilde{\varepsilon})=\rho_0\exp\left[-{\pi\over
        2}e^4\int_{{\sqrt{2}e^2/L}}^{\infty}
      \frac{\rho(\tilde{\varepsilon}')\,d\tilde{\varepsilon}'}
      {(\tilde{\varepsilon}+\tilde{\varepsilon}')^2}\right]}.
\end{equation}
Again, the asymptotic solution is
\begin{equation}
  \label{eq:7}
  {\rho(\tilde{\varepsilon}) = \frac{2}{\pi e^4} \left(
      a{\sqrt{2}\frac{e^2}{L}} +
      |\varepsilon|\right),}
\end{equation}
where $a\approx 0.567$ is the solution of $\ln a = -a$.

\begin{figure}
  \begin{center}
    \epsfysize=3.74cm\epsfbox{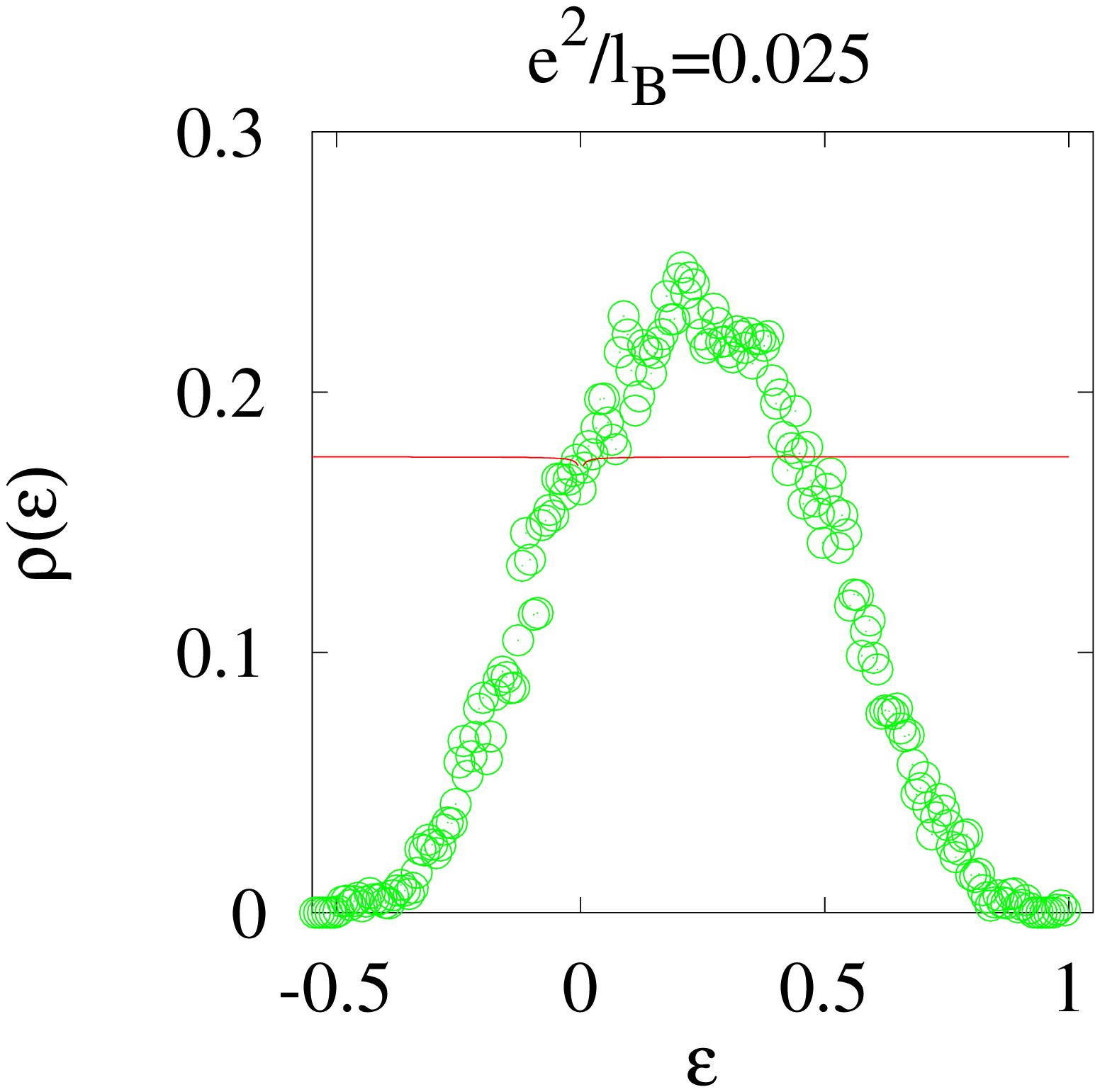}\hspace*{0.4cm}\raisebox{15.2pt}{\epsfysize=3.2cm\epsfbox{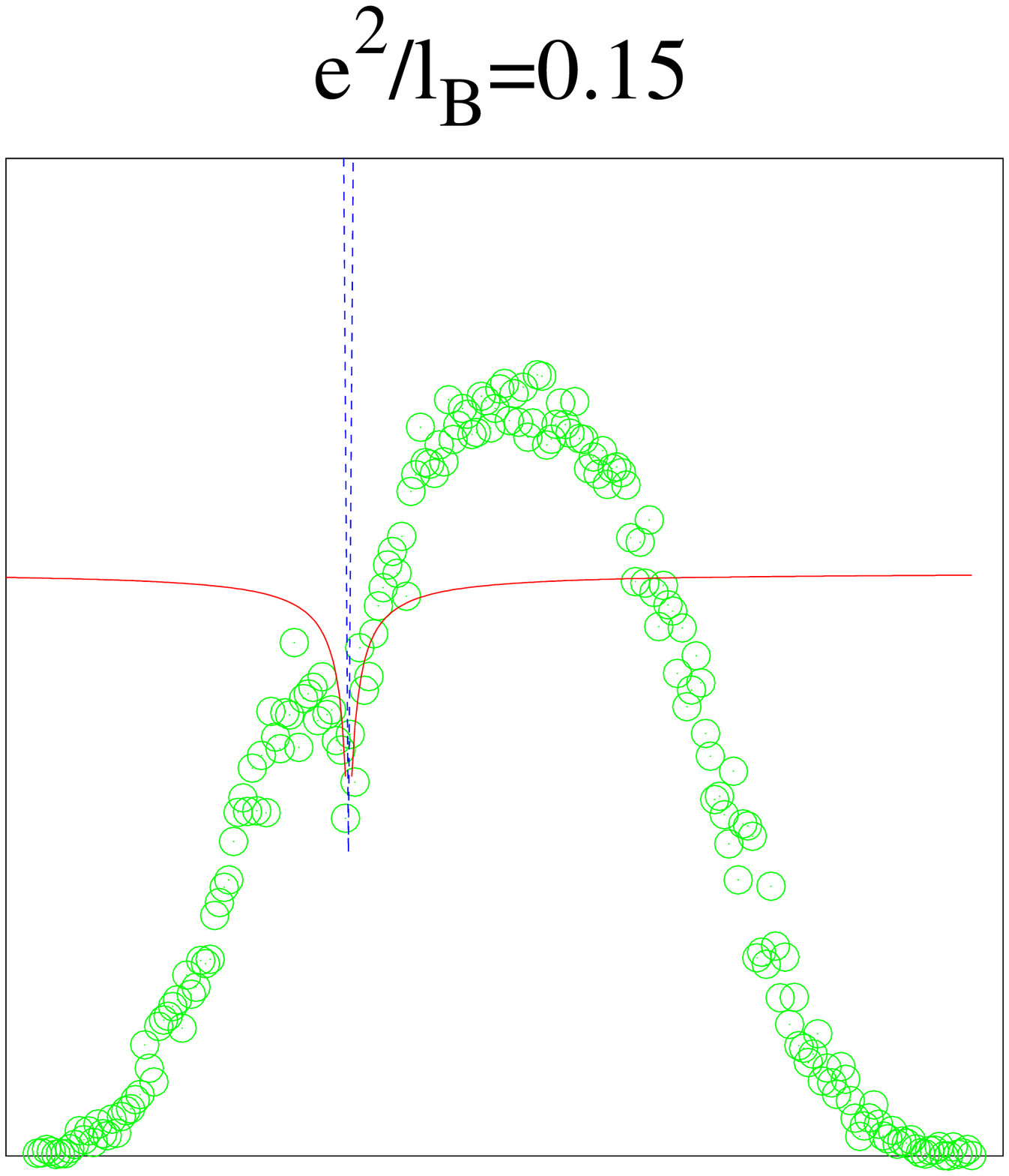}}\hspace*{0.4cm}\raisebox{15.2pt}{\epsfysize=3.2cm\epsfbox{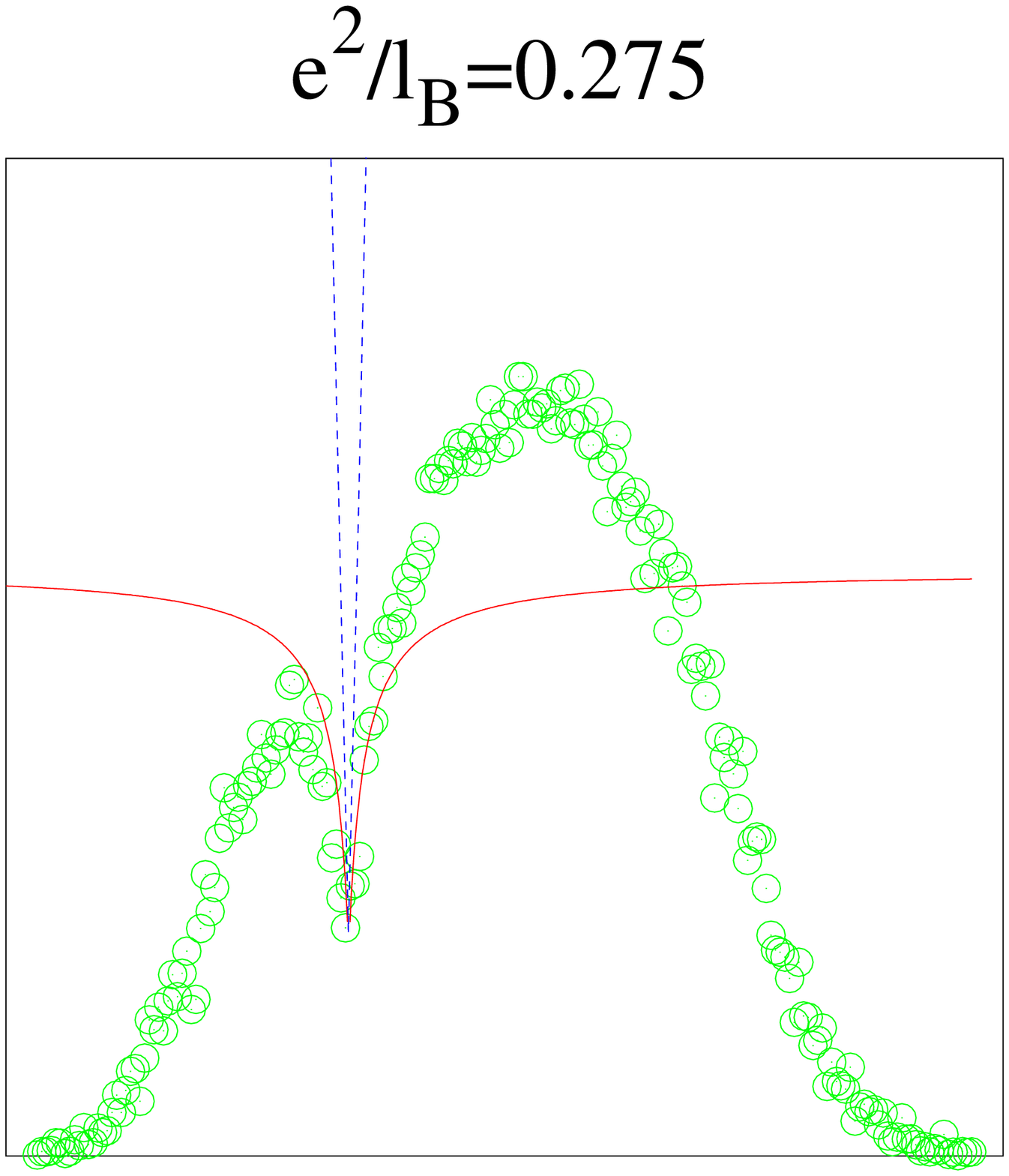}}\\[0.4\baselineskip]
    \hspace*{32pt}\epsfysize=3.2cm\epsfbox{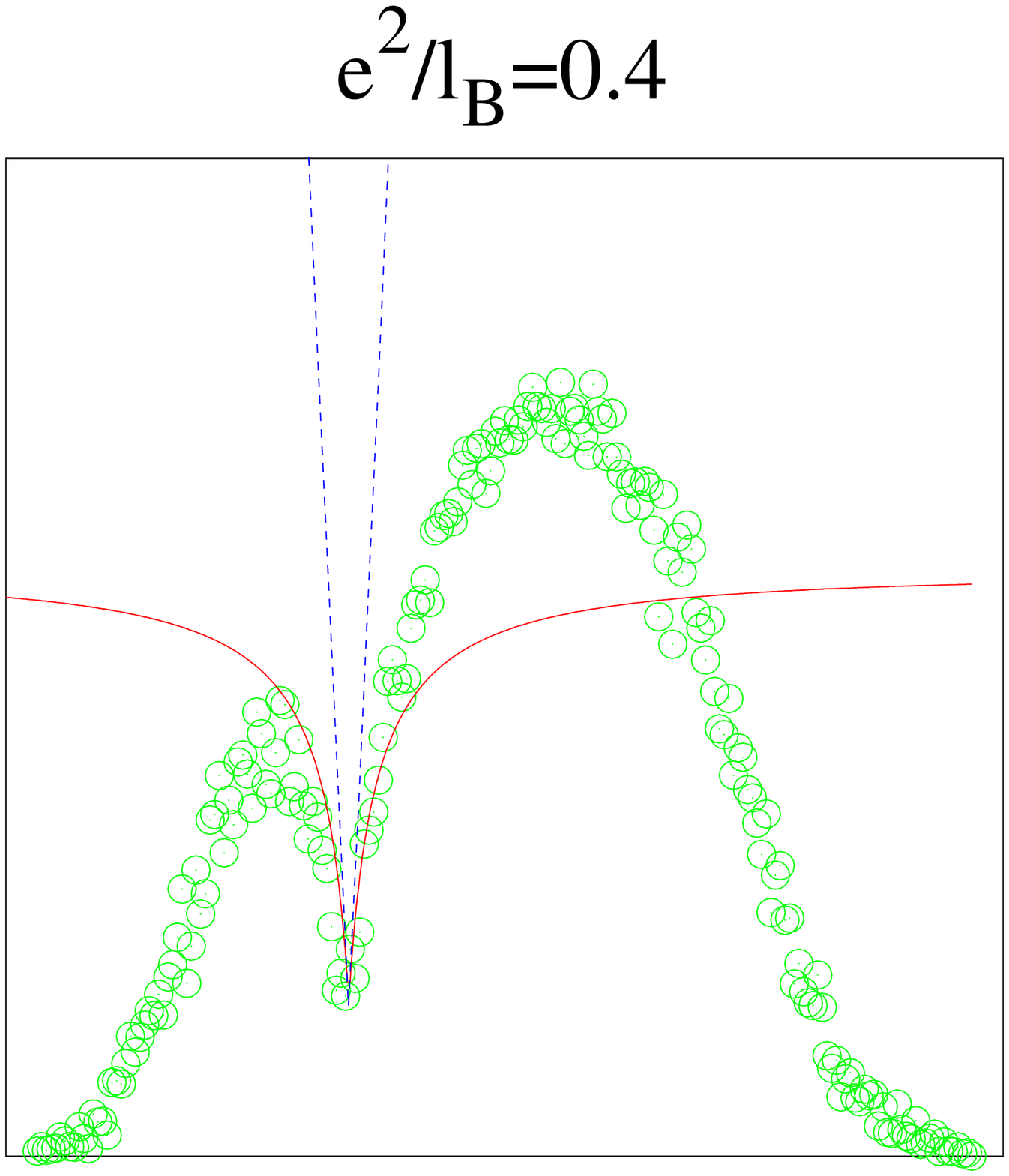}\hspace*{0.4cm}\epsfysize=3.2cm\epsfbox{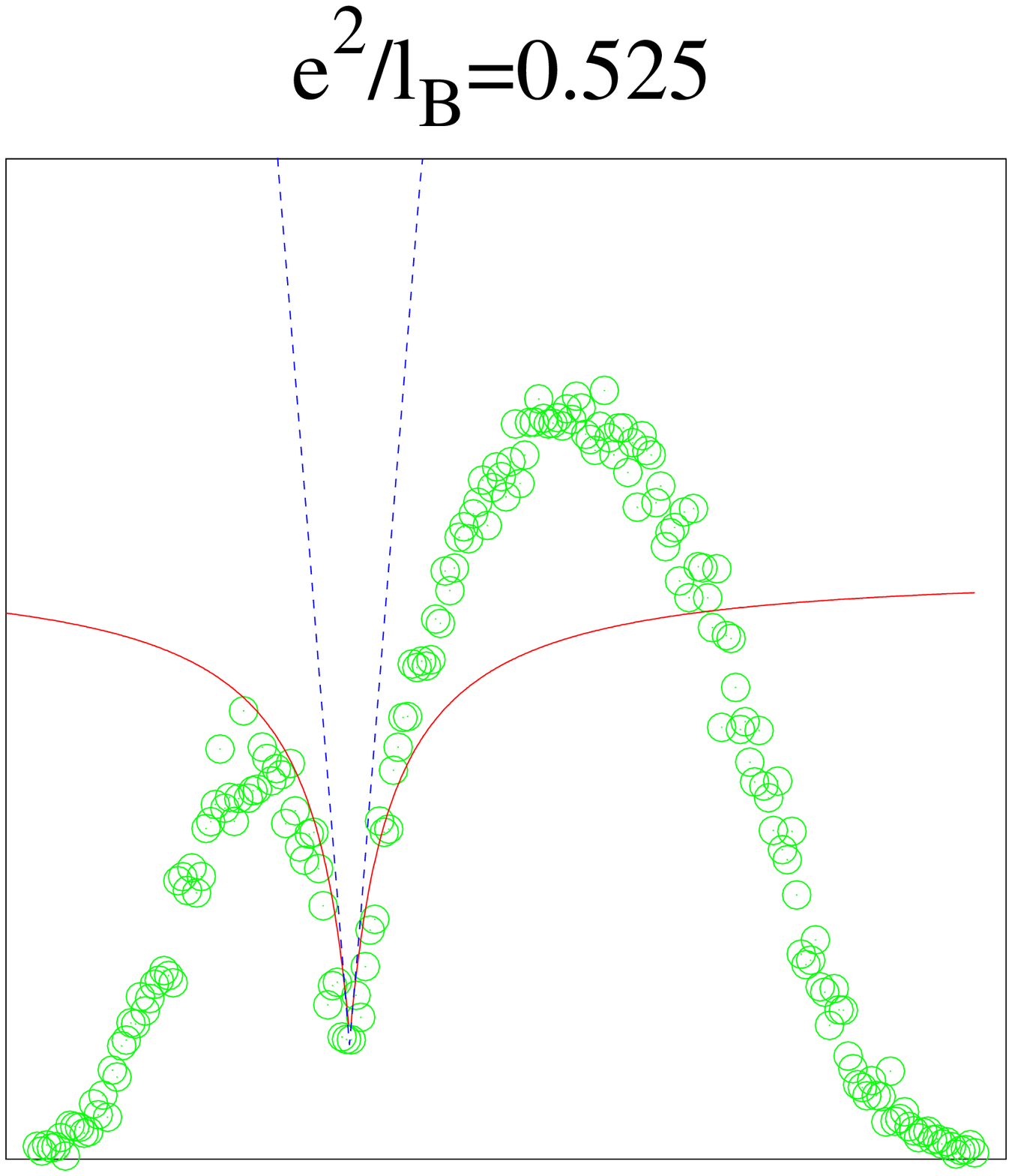}\hspace*{0.4cm}\epsfysize=3.2cm\epsfbox{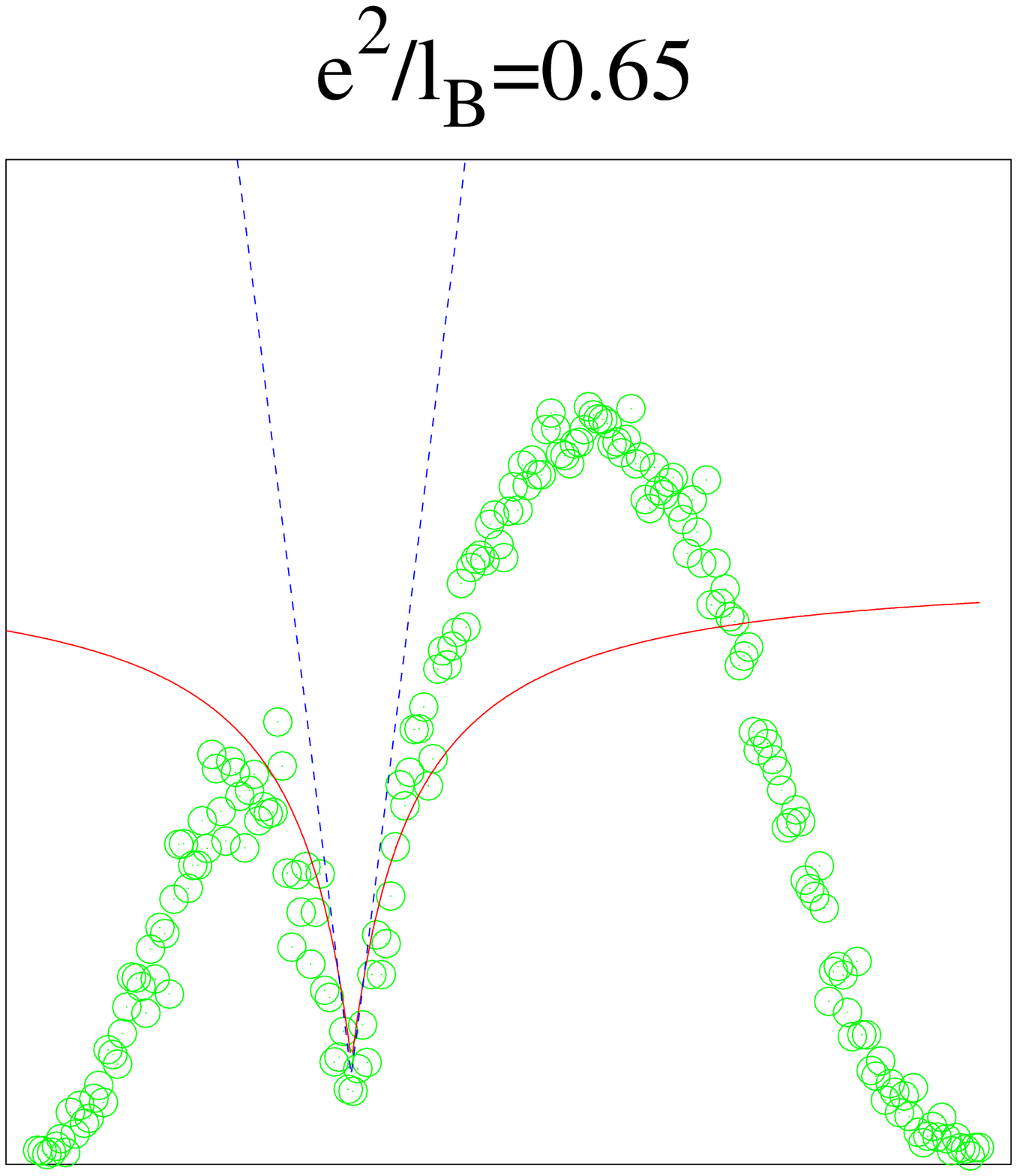}
    \caption{The density of states obtained by smearing every
      eigenvalue by a Gaussian of width $0.8$ mean level spacings for
      interaction strengths $e^2/(l_B\Gamma)$ of 0.025, 0.15, 0.275, 0.4,
      0.525, and 0.65 (top left to right bottom). The dashed lines
      correspond to the solution of eq.~(\ref{eq:4}) and the
      asymptotic solution eq.~(\ref{eq:6}).}
    \label{fig:doss}
  \end{center}
\end{figure}

We solved the self-consistent HF equations for lowest Landau level
with a short-ranged disorder potential \cite{YM93,HB99}. The resulting 
DOS for a system containing 400 flux quanta at filling factor 1/5 
averaged over 60 disorder realizations is shown in
Fig.~\ref{fig:doss}. In order to avoid a filling in of the gap due to
fluctuations of the Fermi energy, the spectra were shifted to assure
$\varepsilon_N+\varepsilon_{N+1}=0$ for every disorder realization,
where $N$ is the number of electrons. Comparison with the
self-consistent solution of eq.~(\ref{eq:4}) and its asymptotic
solution eq.~(\ref{eq:6}) shows qualitative agreement. The deviations 
at larger distance from the Fermi energy are due to the assumption of
a constant DOS in the derivation of eq.~(\ref{eq:4}).

\begin{figure}
  \begin{center}
    \epsfysize=4cm\epsfbox{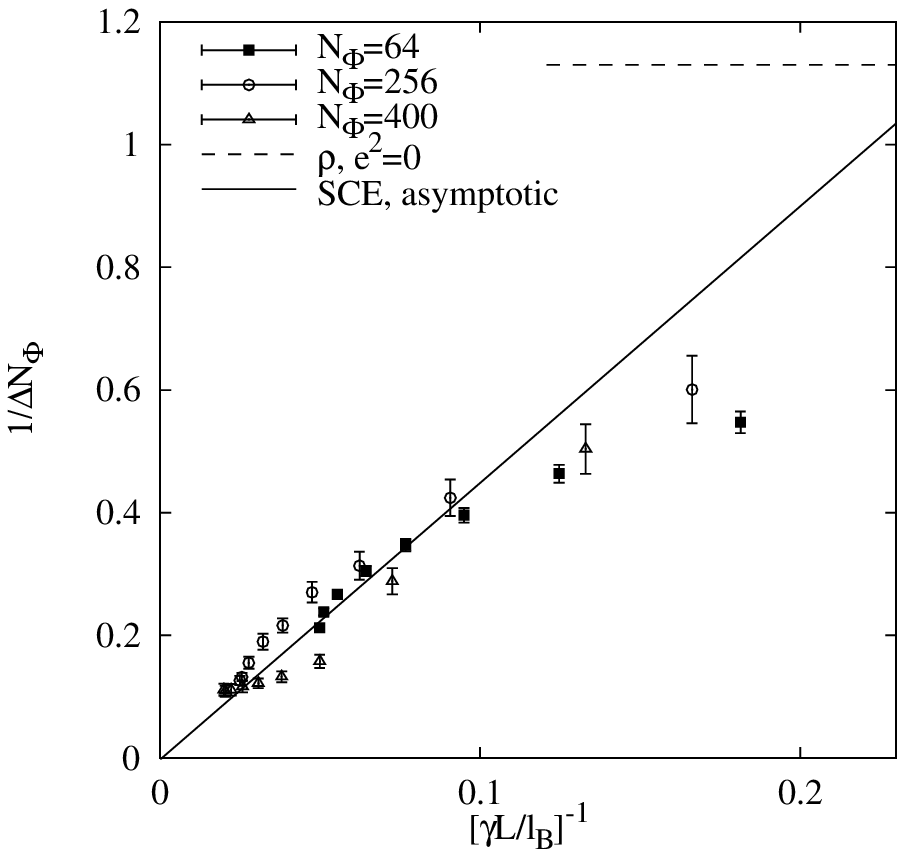}\hspace*{0.5cm}\epsfysize=4.2cm\epsfbox{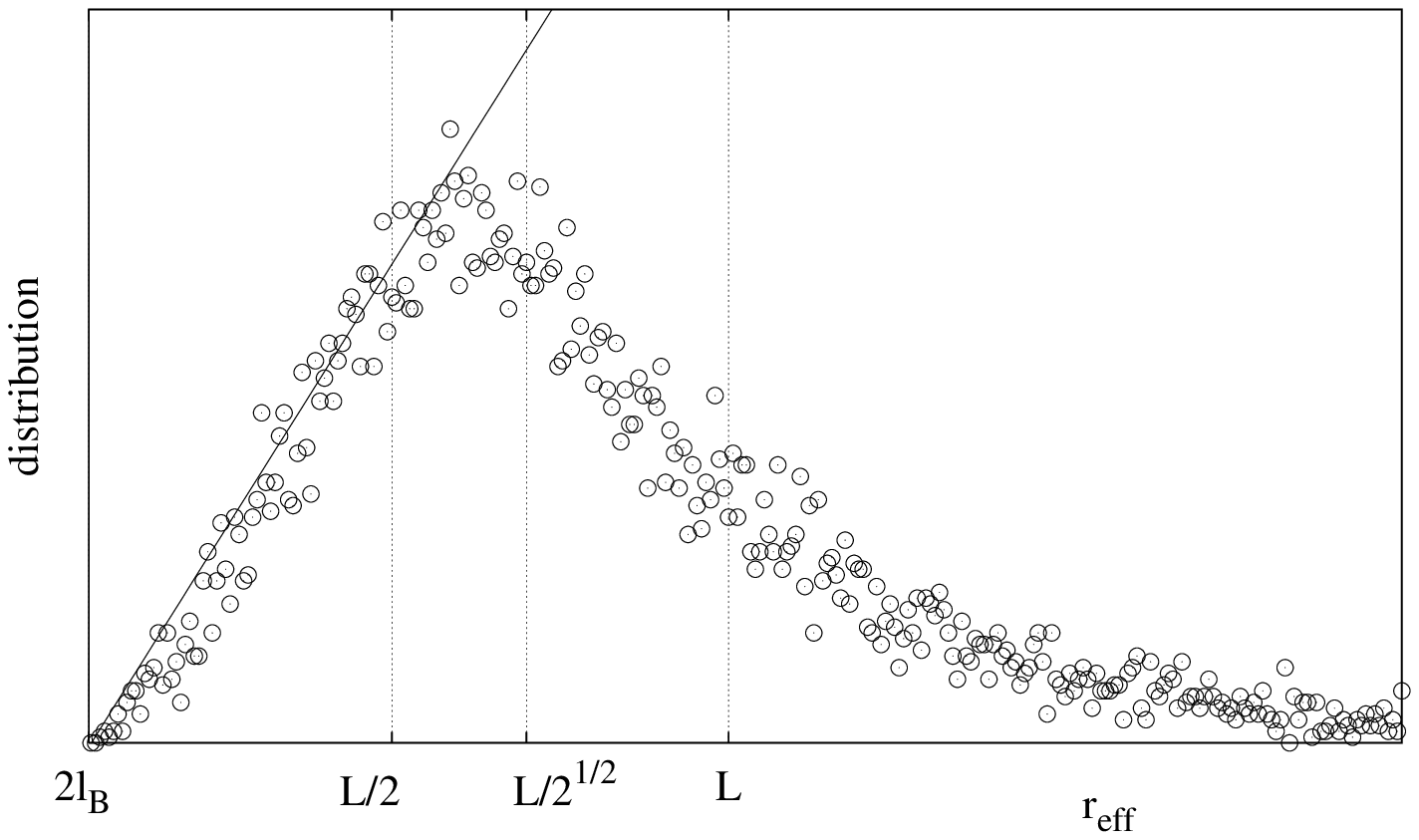}
    \caption{(a) The averaged DOS at the Fermi energy
      $\rho(0)=1/(L^2(\varepsilon_{N+1}-\varepsilon_N))$ of the true
      HF eigenvalues as a function of $l_B/(\gamma L)$. (b)
      Distribution of effective distance $r_{\mathrm{eff}}$ of the HF
      states.}
    \label{fig:dos_true}
  \end{center}
\end{figure}

The DOS of the true HF eigenvalues $\tilde{\varepsilon}$ shows
qualitatively the same behavior. In Fig.~\ref{fig:dos_true} a) the DOS
at the Fermi energy is shown for different system sizes and disorder
strength. The data depend only on $l_B/(\gamma L)$, where
$\gamma=e^2/(l_B\Gamma)$ is a measure of the strength of the
interactions relative to the disorder $\Gamma$, and for small
arguments are well described by the asymptotic solution
eq.~(\ref{eq:7}).

The agreement between numerical simulations and the classical
self-consistent method is surprising, as the conditions for its
applicability are only rather poorly fulfilled in our system. The
localization length $\xi\approx4l_B$ of the non-interacting system is
only slightly smaller than the mean particle spacing of
$(4\pi/(3^{1/2}\nu))^{1/2}l_B\approx6l_B$ \cite{Huc95r}. Thus one
might expect to see stronger effects of the finite wavefunction
overlap and the exchange interaction. Indeed, the distribution of
interaction matrix elements shows these effects. Corresponding to a
matrix element $v_{ij}$ we define an effective distance
$r_{\mathrm{eff}}=e^2/(v_{ij}+v_{0})$ and plot its histogram in
Fig.~\ref{fig:dos_true} b). For a random distribution of classical
point charges one would expect a distribution that rises linearly for
small distances and that terminates at the maximum distance
$L/\sqrt{2}$. In contrast, effective distances much larger than $L$
are present corresponding to states with small interaction matrix
elements due to the exchange interaction.


We studied numerically the Coulomb gap at low filling factor in an
interacting, disordered model of the integer quantum Hall
effect. Surprisingly, we find quantitative agreement with the
predictions of Efros' classical self-consistent method, despite the
fact that the distribution of interaction matrix elements shows strong 
effects of the quantum mechanical effects of wavefunction overlap and
exchange.


This work was performed within the research program of the SFB 341 of
the DFG.

\end{document}